\documentclass[article,nofootinbib]{revtex4-2}

\usepackage{graphicx}
\usepackage{amsfonts}
\usepackage{slashed}
\usepackage{amsmath}
\usepackage{subfigure}
\usepackage{tabularx}
\usepackage{amssymb}
\usepackage{pifont}

\begin{document}

\newcommand{\be}{\begin{equation}}
\newcommand{\ee}{\end{equation}}
\newcommand{\bq}{\begin{eqnarray}}
\newcommand{\eq}{\end{eqnarray}}
\newcommand{\ba}{\begin{align}}
\newcommand{\ea}{\end{align}}

\renewcommand{\a}{\alpha}
\newcommand{\ad}{\dot{\alpha}}
\renewcommand{\b}{\beta}
\newcommand{\bd}{\dot{\beta}}
\newcommand{\q}{\theta}
\newcommand{\pa}{\partial}
\renewcommand{\l}{\lambda}
\newcommand{\g}{\gamma}
\newcommand{\f}{\phi}
\newcommand{\F}{\Phi}
\newcommand{\bD}{\bar{D}}
\newcommand{\tr}{{\rm tr}}
\newcommand{\Str}{{\rm Str}}
\newcommand{\D}{{\cal D}}
\renewcommand{\L}{\Lambda}
\newcommand{\hf}{\frac{1}{2}}
\newcommand{\e}{\epsilon}
\renewcommand{\O}{\Omega}

\newcommand{\cmark}{\ding{51}}
\newcommand{\xmark}{\ding{55}}

\newcommand{\Dslash}{\hbox{$\partial\!\!\!{\slash}$}}
\newcommand{\qslash}{\hbox{$q\!\!\!{\slash}$}}
\newcommand{\pslash}{\hbox{$p\!\!\!{\slash}$}}
\newcommand{\bslash}{\hbox{$b\!\!\!{\slash}$}}
\newcommand{\kslash}{\hbox{$k\!\!\!{\slash}$}}

\title{Scale anomaly in a Lorentz and CPT-violating Quantum Electrodynamics}

\date{\today}

\author{J. S. Porto $^{(a)}$} \email[]{joilson.porto@ifam.edu.br}
\author{A. R. Vieira $^{(b)}$} \email[]{alexandre.vieira@uftm.edu.br}

\affiliation{(a) Instituto Federal do Amazonas, Campus Eirunep\'e, 69880-000, Eirunep\'e, AM, Brasil}
\affiliation{(b) Universidade Federal do Tri\^{a}ngulo Mineiro, Campus Iturama, 38280-000, Iturama, MG, Brasil}

\begin{abstract}
We compute the classical and the quantum breaking of the dilatation current in the minimal Lorentz and CPT-violating quantum electrodynamics. At the classical level, scale 
symmetry is broken by the general mass term $\bar{\psi}M\psi$ and the Chern-Simons-like term. At the quantum level, it is broken by all expected observable fermion operators in 
the massless limit.
\end{abstract}


\maketitle

\section{Introduction}
\label{s1}

Lorentz symmetry is a fundamental property of local relativistic theories and CPT symmetry holds whenever this symmetry is obeyed. To test both of these symmetries, 
the Standard Model Extension (SME) \cite{SME, SME2} proposes all possible terms that break Lorentz and CPT-symmetry that could have emerged at low energies from a 
spontaneous symmetry breaking that occurs at the Planck scale. The QED sector of the SME, known as the QED extension, has been widely tested and the search for Lorentz and CPT 
violation via this sector reveals how tiny the violation of these symmetries would be. It suffices to quote that tests on the electron \cite{electrons0}-\cite{electrons3} or the pure 
photon sectors \cite{photon1}-\cite{photon4} show how good these symmetries are. 

Symmetries like the Lorentz and CPT ones are one of the the main criteria in the construction of a theory, among others like gauge symmetry. It was at first thought that if 
theories possess symmetries at the classical level, they would keep these symmetries at the quantum level. But this is a naive assumption after such theory is quantized. The 
breaking of classical symmetries when quantum corrections are taken into account are called anomalies. Some of them are measurable because they are related to physical processes, 
like the Adler-Bell-Jackiw anomaly (ABJ-anomaly) \cite{Bell, Adler}, related to the pion decay into two photons, and the trace anomaly, related to hadronic processes \cite
{Ellis}. The trace of the energy-momentum tensor, which is equal to the dilatation current conservation associated to scale symmetry, is broken at the classical level by 
massive terms so that massless field theories are scale invariant. However, using these theories for short distances reveals that scale symmetry is broken when loop corrections 
are taken into account \cite{Callan}.

The trace anomaly also manifests itself as a breaking of conformal symmetry in curved space when matter fields are embedded in a gravitational background \cite{Duff}. 
The computation performed with different regularization schemes reveals universal pieces of this symmetry breaking in the anomalous energy-momentum trace \cite{Duff2}-\cite
{IRegTA}. From physics point of view, it is interesting to know if the anomaly is an artifact of the regularization employed or if it is indeed physical since a regularization 
scheme can spuriously break symmetries of the theory. This happens in particular when the Lagrangian of the theory has dimensional specific objects, like $\gamma_5$ matrices and 
Levi-Civita symbols, like some of the Lorentz and CPT-violating operators of the QED extension.  In the usual quantum electrodynamics, the anomalous breaking of this 
energy-momentum tensor was found for different regularization schemes \cite{Ellis}-\cite{LoopReg} and it is worthy computing this anomaly in QED extension.

In this work, we compute the breaking of the dilatation current at one-loop order in a Lorentz and CPT- violating Quantum Electrodynamics. In particular, it is considered the minimal sector
of the SME, where only terms renormalizable by power counting are included in the Lagrangian. In section \ref{s2}, we present an overview on the scale symmetry of this model and its possible anomalous 
amplitudes. In section \ref{s3}, we present a summary of the regularization scheme employed. In section \ref{s4}, we perform the one-loop calculation and we present conclusions in section \ref{s5}.

\section{Scale symmetry and dilatation current in QED extension}
\label{s2}

A theory is scale invariant if it is unchanged under the following symmetry transformations,

\be
x'= e^{\epsilon} x
\label{eq0}
\ee

and

\be
\Phi'(x')= e^{-\epsilon d_{\Phi}} \Phi(x),
\label{eq1}
\ee
where $\epsilon$ is a scale parameter, $\Phi$ is a generic field and $d_{\Phi}$ is its scale dimension.

As an example, the kinetic terms of the electromagnetic or spinor fields, $-\frac{1}{4}F^{\mu\nu}F_{\mu\nu}$ and $\bar{\psi}\gamma^{\mu}\partial_{\mu}\psi$, respectively, are scale
invariant because they possess scale dimension equal to 4. In this way, when the volume element $d^4x$ changes in the action, it compensates the change in the fields making
the theory scale invariant. On the other hand, mass terms or interaction terms with dimensional coupling constants, break scale symmetry at the classical level.

The Noether current related to the transformations of eqs. (\ref{eq0}) and (\ref{eq1}) is the dilatation current $\mathcal{J}^{\mu}=\Pi^{\mu}d_{\Phi}\Phi+x_{\nu}\Theta^{\mu\nu}$, 
where $\Pi^{\mu}=\frac{\partial \mathcal{L}}{\partial (\partial_{\mu}\Phi)}$ is the conjugate momentum and $\Theta^{\mu\nu}$ is the canonical energy momentum tensor given by 
$\Theta^{\mu\nu}=\Pi^{\mu}\partial^{\nu}\Phi-\eta_{\mu\nu}\mathcal{L}$.

Besides the dimensional terms, like mass terms in the lagrangian, the renormalization process also breaks scale symmetry because it introduces a renormalization group scale
dependence in the couplings. This is known as the anomalous breaking of the scale symmetry.

It is possible to study this scale symmetry breaking in a Lorentz and CPT-violating version of QED. We consider the fermion sector of the following 
lagrangian

\be
\mathcal{L} =\frac{1}{2}i \bar{\psi}\Gamma^{\mu}\overleftrightarrow{D}_{\mu}\psi- \bar{\psi}M\psi-\frac{1}{4}F^{\mu\nu}F_{\mu\nu}
-\frac{1}{4}(k_F)_{\kappa\lambda\mu\nu}F^{\mu\nu}F^{\kappa\lambda}+ \frac{1}{2}(k_{AF})^{\kappa}\epsilon_{\kappa\lambda\mu\nu}A^{\lambda}F^{\mu\nu},
\label{EQ1}
\ee
where $D_{\mu}\equiv \partial_{\mu}+ieA_{\mu}$ is the usual covariant derivative which couples the gauge field with matter,

\begin{align}
&\Gamma^{\nu}=\gamma^{\nu}+\Gamma^{\nu}_1,\nonumber\\
&\Gamma^{\nu}_1= c^{\mu\nu}\gamma_{\mu}+d^{\mu\nu}\gamma_5\gamma_{\mu}+e^{\nu}+i f^{\nu}\gamma_5+\frac{1}{2}\eta_{\lambda\mu\nu}
\sigma_{\lambda\mu}
\end{align}
and
\begin{align}
&M=m+M_1, \nonumber\\
&M_1= im_5 \gamma_5 +a^{\mu}\gamma_{\mu}+b_{\mu}\gamma_5\gamma^{\mu}+\frac{1}{2}H_{\mu\nu}\sigma^{\mu\nu}.
\end{align}

The coefficients $a_{\mu}$, $b_{\mu}$, $c_{\mu\nu}$, $d_{\mu\nu}$, $e_{\mu}$, $f_{\mu}$, $g_{\lambda\mu\nu}$, $H_{\mu\nu}$, 
$(k_F)_{\kappa\lambda\mu\nu}$ and $(k_{AF})_{\kappa}$ govern Lorentz violation and only the coefficients $a_{\mu}$, $b_{\mu}$, $e_{\mu}$, $f_{\mu}$, $g_{\lambda\mu\nu}$ and 
$(k_{AF})_{\kappa}$ govern CPT violation since the number of indices is odd. Not all of these coefficients are expected to be observable since some combinations of them
can be removed by a field redefinition \cite{Neil}. There is also a  subtleness regarding the $f_{\mu}$ coefficients because, unlike $a_{\mu}$ and $e_{\mu}$, its time component
is odd under all spacial reflection \cite{Bretf}. 

We expect that scale symmetry be broken at the classical level by all the dimensional coefficients of the lagrangian (\ref{EQ1}). We see this would be the case of the 
Chern-Simons-like term $(k_{AF})^{\kappa}\epsilon_{\kappa\lambda\mu\nu}A^{\lambda}F^{\mu\nu}$ or the generalized mass term $\bar{\psi}M\psi$.

As in the usual electrodynamics, it is possible to build a symmetric energy-momentum tensor, known as the Belinfante tensor whose trace
leads to the classical break of the dilatation current. We explicitly show in the appendix \ref{A1} that the break of the dilatation current in the classical theory
is given by
\be
\partial_{\mu}\mathcal{J}^{\mu}=T^{\mu}_{\mu}=\bar{\psi}M\psi,
\ee
where $T^{\mu}_{\mu}$ is the trace of the symmetric energy-momentum tensor.

The quantum corrections also break the dilatation current. Applying naively the scale transformations of eqs. (\ref{eq0}) and (\ref{eq1}), we find that the Ward identity of this dilatation current is given by
\be
-i G_T(0,p_1,p_2,...,p_{n-1})=\left(n(d_{\Phi}-4)+4-\sum_k^{n-1}p_k\frac{\partial}{\partial p_k}\right)G(p_1,p_2,...,p_{n-1}),
\label{eqIW}
\ee
where $G_T(0,p_1,p_2,...,p_{n-1})$ and $G(p_1,p_2,...,p_{n-1})$ are $(n-1)$-point Green functions in the momentum space of the respective
coordinate space $n-$point Green functions, $G(y,x_1,x_2,...,x_{n})=\langle T^{\mu}_{\mu}(y)\Phi(x_1)\Phi(x_2)...\Phi(x_{n})\rangle$ and $G(x_1,x_2,...,x_{n})=\langle 
\Phi(x_1)\Phi(x_2)...\Phi(x_{n})\rangle$. Note that the lower index $T$ is to emphasize that one of the fields in the expectation value is the energy-momentum tensor. Besides,
the derivation of eq. (\ref{eqIW}) assumes $\mathcal{D}\Phi=\mathcal{D}\Phi'$ in a scale transformation of the functional integral. Since this is not true for quantum corrections,
we are going to see in what follows that this equation is incomplete, {\it i.e.} there is an additional anomalous term on it.

For $n=2$ and $\Phi$ being the vector field, eq. (\ref{eqIW}) reads
\bq
&-i G^{\mu\nu}_T(0,p)=\left(n(d_{\Phi}-4)+4-p^{\lambda}\frac{\partial}{\partial p^{\lambda}}\right)G^{\mu\nu}(p), \nonumber\\
&-i G^{\mu\nu}_T(0,p)=\left(-2-p^{\lambda}\frac{\partial}{\partial p^{\lambda}}\right)G^{\mu\nu}(p).
\label{eqIWG}
\eq

It is also possible to rewrite (\ref{eqIWG}) in terms of the one-particle-irreducible (1PI) function $\Gamma^{\sigma\lambda}(p)$, related with the legged 2-point function
via the equation $G_{\mu\nu}(p)=D_{\mu\sigma}(p)\Gamma^{\sigma\lambda}(p)D_{\lambda\nu}(p)$, being $D_{\mu\nu}(p)=-\frac{i\eta_{\mu\nu}}{p^2}$ the photon propagator. The Ward
identity in eq. (\ref{eqIWG}) can be written as

\be
-i \Gamma^{\mu\nu}_T(0,p)=\left(2-p^{\lambda}\frac{\partial}{\partial p^{\lambda}}\right)\Gamma^{\mu\nu}(p).
\ee

\begin{figure}
\centering
\subfigure[]{\includegraphics[trim=0mm 65mm 0mm 65mm,scale=0.45]{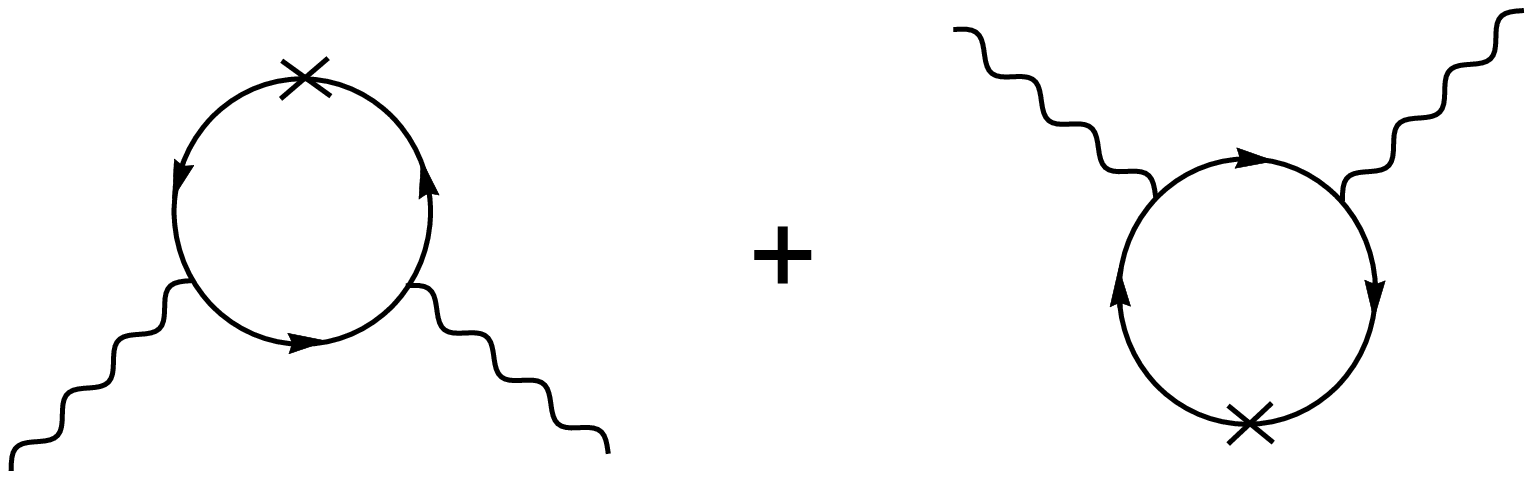}}
\subfigure[]{\includegraphics[trim=0mm 15mm 0mm 15mm,scale=0.45]{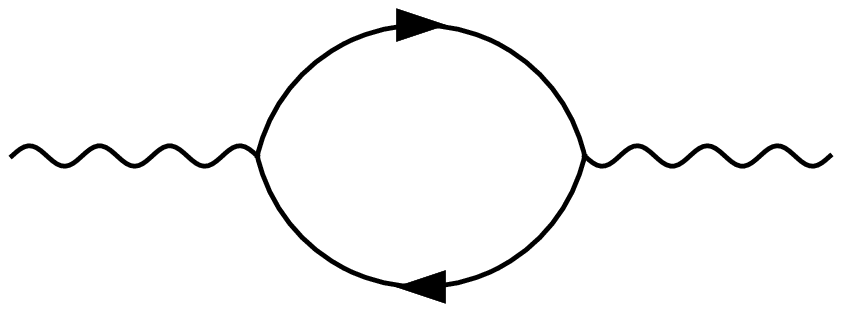}}
\caption{(a) One-loop 3-point Feynman diagram with the energy-momentum tensor as a vertex represented by a cross. (b) Vacuum polarization tensor.}
\label{fig1}
\end{figure}

Finally, $\Gamma^{\mu\nu}_T(0,p)$ corresponds to the 3-point Feynman diagram $\Delta^{\mu\nu}(0,p,-p)$ shown in figure \ref{fig1}, where two of the vertices are the usual ones
and the other is the trace of the energy-momentum tensor $T^{\mu}_{\mu}=\bar{\psi}M\psi$,  $\Gamma^{\mu\nu}(p)$ corresponds to the 2-point Feynman diagram $\Pi^{\mu\nu}(p,-p)$, 
which is the vacuum polarization tensor. In terms of these diagrams the Ward identity can be written as:
\be
\Delta^{\mu\nu}=\left(2-p^{\lambda}\frac{\partial}{\partial p^{\lambda}}\right)\Pi^{\mu\nu}(p).
\label{eqIWD}
\ee

According to lagrangian (\ref{EQ1}), the solid line in the amplitudes of figure \ref{fig1} is the full fermion propagator $S(k)=\frac{i}{\Gamma_{\mu}k^{\mu}-M}$ and the 
electron-photon interaction is $-ie\Gamma_{\mu}$. Thus, besides the usual electro-dynamical contribution, we have first order in the Lorentz and CPT-violating coefficients.

Since one-loop diagrams require a regularization and a renormalization procedure, a scale is introduced in this process and computing the diagrams of both sides of
eq. (\ref{eqIWD}) yields the quantum breaking of scale symmetry expressed by the non-zero trace of the energy-momentum tensor.

\section{Implicit Regularization}
\label{s3}

We apply the implicit regularization (IReg) framework \cite{Orimar}-\cite{Luciana} to treat the divergent integrals of the amplitudes of the following. In this scheme, 
an implicit regulator $\Lambda$, not a hard cutoff necessarily, is assumed so that it makes sense to perform algebraic operations with the integrands. It is possible to use,
for example, the following identity to separate UV basic divergent integrals (BDI) from the finite part:

\be
\int_k \frac{1}{(k+p)^2-m^2}=\int_k\frac{1}{k^2-m^2}-\int_k\frac{(p^2+2p\cdot k)}{(k^2-m^2)[(k+p)^2-m^2]},
\label{2.1}
\ee
where $\int_k\equiv\int^\Lambda\frac{d^4 k}{(2\pi)^4}$. These BDI's are defined as follows:
\begin{equation}
I^{\mu_1 \cdots \mu_{2n}}_{log}(m^2)\equiv \int_k \frac{k^{\mu_1}\cdots k^{\mu_{2n}}}{(k^2-m^2)^{2+n}},
\end{equation}
and
\begin{equation}
I^{\mu_1 \cdots \mu_{2n}}_{quad}(m^2)\equiv \int_k \frac{k^{\mu_1}\cdots k^{\mu_{2n}}}{(k^2-m^2)^{1+n}}.
\end{equation}

The basic divergences with Lorentz indices can be combined as differences between integrals with the same superficial degree 
of divergence, according to the equations below, which define surface terms \footnote{The Lorentz indices between brackets stand for 
permutations, i.e. $A^{\{\alpha_1\cdots\alpha_n\}}B^{\{\beta_1\cdots\beta_n\}}=A^{\alpha_1\cdots\alpha_{n}}B^{\beta_1\cdots\beta_n}$ 
+ sum over permutations between the two sets of indices $\alpha_1\cdots\alpha_{n}$ and $\beta_1\cdots\beta_n$}:
\begin{align}
&\Upsilon^{\mu \nu}_{2w}= \eta^{\mu \nu}I_{2w}(m^2)-2(2-w)I^{\mu \nu}_{2w}(m^2) 
\equiv \upsilon_{2w}\eta^{\mu \nu},
\label{dif1}\\
\nonumber\\
&\Xi^{\mu \nu \alpha \beta}_{2w}=  \eta^{\{ \mu \nu} \eta^{ \alpha \beta \}}I_{2w}(m^2)
 -\nonumber\\&- 4(3-w)(2-w)I^{\mu \nu \alpha \beta }_{2w}(m^2) \equiv  \xi_{2w}\eta^{\{ \mu \nu} \eta^{ \alpha \beta \}},
\label{dif2}\\
\nonumber\\
&\Sigma^{\mu \nu \alpha \beta \gamma \delta}_{2w} = g^{\{\mu \nu} g^{ \alpha \beta} g^ {\gamma \delta \}}I_{2w}(m^2)
-8(4-w)(3-w)(2-w) I^{\mu \nu \alpha \beta \gamma \delta}_{2w}(m^2)= \sigma_{2w} g^{\{\mu \nu} g^{ \alpha \beta} g^ {\gamma \delta \}}.
\label{dif3}
\end{align}

In the expressions above, $2w$ is the degree of divergence of the integrals and  for the sake of brevity, we substitute the subscripts 
$log$ and $quad$ by $0$ and $2$, respectively. Surface terms can be conveniently written as integrals of total 
derivatives, namely
\be
\upsilon_{2w}\eta^{\mu \nu}= \int_k\frac{\partial}{\partial k_{\nu}}\frac{k^{\mu}}{(k^2-m^2)^{2-w}},
\label{ts1}
\ee
\be
(\xi_{2w}-v_{2w})\eta^{\{ \mu \nu} \eta^{ \alpha \beta \}}= \int_k\frac{\partial}{\partial k_{\nu}}\frac{2(2-w)k^{\mu} k^{\alpha} k^{
\beta}}{(k^2-m^2)^{3-w}},
\label{ts2}
\ee
and
\be
(\sigma_{2w}-\xi_{2w})\eta^{\{ \mu \nu} \eta^{ \alpha \beta} \eta^ {\gamma \delta \}}=\int_k\frac{\partial}{\partial k_{\nu}}\frac{4(3-w)(2-w)k^{\mu} k^{\alpha} k^{\beta} k^{\gamma} k^{\delta}}{(k^2-m^2)^{4-w}}.\\
\label{ts3}
\ee

Equation (\ref{2.1}) is applied in a recursive way till the BDI's is separated from the finite content. Also, we see that equations (\ref{dif1})-(\ref{dif3}) are
undetermined because they are differences between divergent quantities. Each regularization scheme gives a different value for these terms. However, as physics should
not depend on the regularization, we leave these terms to be arbitrary until the end of the calculation, fixing them by symmetry constraints, like a Ward identity \cite{Jackiw2}.


The prescription above is not the usual regularization procedure and it is particularly useful in the computation of
anomalies, where we need to know if the quantum breaking of a symmetry really occurs or if it is a spurious symmetry breaking caused
by the regulator.



\section{One-loop breaking of the dilatation current}
\label{s4}

The amplitudes of the diagrams in figure \ref{fig1} can be easily constructed using the Feynman rules, as we see below:
\bq
&\Delta_{\mu\nu}=-2\int \frac{d^4k}{(2\pi )^4} Tr[e \Gamma_{\mu}S_{LV}(k)e \Gamma_{\nu}S_{LV}(k+p)M S_{LV}(k+p)],
\nonumber\\
&i\Pi^{\mu\nu}= \int \frac{d^4k}{(2\pi )^4} Tr[e \Gamma_{\mu}S_{LV}(k)e \Gamma_{\nu} S_{LV}(k+p)],
\label{Amps}
\eq
where $S_{LV}(k)$ is the fermion propagator modified by Lorentz and CPT violation. It can be expanded in a series where we neglect the terms beyond first order in Lorentz and
CPT violation according to the equation below:
\be
\frac{i}{\kslash+\Gamma^{\mu}_1k_{\mu}-M}=\sum_{n=0}^{\infty} \frac{i}{\kslash-M}\left(i\Gamma^{\mu}_1k_{\mu}\frac{i}{\kslash-M}\right)^n,
\label{series}
\ee
where the same expansion can be done considering the general mass term $M$.

We list all finite and regularized divergent integrals in appendix \ref{A2}. For instance, the result of the amplitudes in eq. (\ref{Amps}) for the isotropic case and
for the $b_{\mu}$ coefficient is given by
\bq
&\Delta_{\mu\nu}=\frac{m^2 e^2}{2\pi^2}(\eta_{\mu\nu}Z_0+\iota_0 \eta_{\mu\nu}p^2-2\iota_1 p_{\mu}p_{\nu}-16\pi^2 i\upsilon_0 \eta_{\mu\nu})+\nonumber\\
&2e^2\left(8\upsilon_0+\frac{i}{6\pi^2}+\frac{i}{4\pi^2}p^2\int_0^1 dx \frac{x(1-x)}{\D(x)}+\frac{i m^2}{\pi^2}\int_0^1 dx \frac{x^2}{\D(x)}\right)
\epsilon_{\mu\nu\alpha\beta}b^{\alpha}p^{\beta},
\label{eqdelta}
\eq
and
\bq
\centering
i\Pi^{\mu\nu}&=\frac{4}{3}(p^2\eta^{\mu\nu}-p^{\mu}p^{\nu})I_{log}(m^2)-4\upsilon_2\eta^{\mu\nu}+\frac{4}{3}(p^2\eta^{\mu\nu}-p^{\mu}p^{\nu})\upsilon_0+\nonumber\\
&-\frac{4}{3}(p^2\eta^{\mu\nu}+2p^{\mu}p^{\nu})(\xi_0-2\upsilon_0)-\frac{i}{2\pi^2}(p^2\eta^{\mu\nu}-p^{\mu}p^{\nu})(Z_1-Z_2)+\nonumber\\
&+e^2\left(-\frac{m^2}{\pi^2}\iota_0+ 4i\upsilon_0 \right)\epsilon_{\mu\nu\alpha\beta}b^{\alpha}p^{\beta},
\label{eqpi}
\eq
where $Z_n=\int_0^1 dx x^n\ln \left(\frac{\D(x)}{m^2}\right)$, $\iota_n=\int_0^1 dx \frac{x^n(1-x)}{\D(x)}$ and $\D(x)=m^2-p^2 x(1-x)$.

Considering the isotropic case ($b=0$), we see that the quadratic surface term $\upsilon_2$ breaks gauge symmetry and should be set to zero. On the other hand, the logarithmic 
surface term $\upsilon_0$ has a transverse piece, as we can see requiring gauge symmetry $p_{\mu}\Pi^{\mu\nu}=0$. If $\xi_0=0$ and $\upsilon_0=0$, we have gauge symmetry but also 
if $\xi_0=2\upsilon_0$. Thus, it is possible to let the result depending on $\upsilon_0$ and the arbitrariness expressed by this surface term means the freedom of choosing the 
renormalization point. However, there is an unavoidable arbitrariness in the induced Chern-Simons-like term that also can not be fixed requiring gauge symmetry because this term 
is already gauge invariant. That is why the induced CS-like term is regularization dependent, although being finite \cite{CSI}-\cite{CSI11}. In what follows, we adopt all surface 
terms equal to zero. This is compatible with gauge invariance and the choice of a renormalization point.

The renormalization group scale $\lambda$ is introduced with the use of the scale relation below:
\be
I_{log}(m^2)=I_{log}(\lambda^2)+\tilde{b} \ln \left(\frac{\lambda^2}{m^2}\right).
\ee
When the massless limit is taken, $\lambda$ is the only mass scale remaining.

The amplitude $\Delta_{\mu\nu}$ in eqs. (\ref{Amps}) is zero in the massless limit for all terms of the general vertex $-ie\Gamma^{\mu}$. So, all the anomalous terms related to
these coefficients comes from the amplitude $\Pi_{\mu\nu}$. For all coefficients of the fermion sector, inserting eq. (\ref{Amps}) in the Ward identity and taking the massless 
limit to try to restore the conservation of the dilatation current, we find out the anomalous term
\bq
&\Delta^{\mu\nu}_r-\left(2-p^{\lambda}\frac{\partial}{\partial p^{\lambda}}\right)\Pi^{\mu\nu}_r= \frac{e^2}{6\pi^2}(p^2\eta^{\mu\nu}-p^{\mu}p^{\nu})-\frac{i}{6\pi^2}e^2\epsilon^{
\mu\nu\alpha\beta}b_{\alpha}p_{\beta}
-\frac{e^2}{16\pi^2}\frac{c^{pp}}{p^2}(p^2\eta^{\mu\nu}-p^{\mu}p^{\nu})-\nonumber\\
&-\frac{e^2}{6\pi^2}((c^{\nu p}+c^{p \nu})p^{\mu}+(c^{\mu p}+c^{p \mu})p^{\nu}-(c^{\mu \nu}+c^{\nu \mu})p^{2}-2c^{p p}\eta^{\mu \nu}),
\label{result}
\eq
where the index $r$ in the tensors stands for the renormalized amplitudes and $c^{\mu p}\equiv c^{\mu\nu}p_{\nu}$. The isotropic term is compatible with the ones found with other regularization schemes \cite{Ellis}-\cite{LoopReg}. As expected, only observable combinations of the fermion coefficients are found \cite{Neil}. It might appear that the $\Delta^{\mu\nu}$ amplitude
in figure \ref{fig1}a gives a non-zero contribution proportional to the $a_{\mu}$ coefficient. However, this would be equivalent to doing the shift $p\rightarrow p-a$ in the isotropic
amplitude $\Pi^{\mu\nu}(p)$, meaning that there is no additional Lorentz and CPT-violating information coming from the $a_{\mu}$ coefficient in observables \cite{Neil}. 

Table \ref{tab1} summarizes how each Lorentz and CPT-violating coefficient of the fermion sector breaks scale symmetry in the massless limit both in the classical and quantum levels.

\begin{center}
 \begin{tabular}{|c|c|c|c|}
  \hline
  Lagrangian term &Scale Dimension& Classical Breaking & Quantum Breaking \\
  \hline
  $\bar{\psi}a_{\mu}\gamma^{\mu}\psi$ & 3& \cmark & \xmark \\
  \hline
   $\bar{\psi}b_{\mu}\gamma_5\gamma^{\mu}\psi$& 3& \cmark & \cmark \\
  \hline
  $c_{\mu \nu}\bar{\psi}\gamma^{\mu}D^{\nu}\psi$& 4&\xmark & \cmark \\
  \hline
  $d_{\mu\nu}\bar{\psi}\gamma_5\gamma^{\mu}D^{\nu}\psi$& 4&\xmark & \xmark \\
  \hline
  $e_{\mu}\bar{\psi}D^{\mu}\psi$& 4&\xmark & \xmark \\
  \hline
  $f_{\mu}\bar{\psi}\gamma_5D^{\mu}\psi$& 4&\xmark &\xmark \\
  \hline
  $g_{\mu\nu\lambda}\bar{\psi}\sigma^{\mu\nu}D^{\lambda}\psi$& 4&\xmark & \xmark \\
  \hline
  $H_{\mu\nu}\sigma^{\mu\nu}\psi$& 3&\cmark &\xmark \\
  \hline
  $\bar{\psi}m_5\gamma_5\psi$& 3& \cmark & \xmark \\
  \hline 
 \end{tabular}
\label{tab1} 
\end{center}

Eq. (\ref{result}) can be written in terms of fields and it leads to the following trace of the energy-momentum tensor together with the classical breaking:
\be
T^{\mu}_{\mu}=\bar{\psi}M\psi+(k_{AF})_{\mu}A_{\alpha}\tilde{F}^{\alpha\mu}+\frac{\tilde{e}^2}{24\pi^2}F^{\mu\nu}F_{\mu\nu}-\frac{ie^2}{12\pi^2} \epsilon_{\mu\nu\alpha\beta}b^{\alpha} A^{\mu}F^{\beta\nu}+\frac{e^2}{12\pi^2}\bar{c}_{\mu\nu}F^{\mu\alpha}F^{\nu}_{\alpha},
\ee 
where $\tilde{e}$ is the QED coupling with a small LV correction already observed in other contexts \cite{Drummond,Sherrill}, $\bar{c}_{\mu\nu}\equiv c_{\mu\nu}+c_{\nu\mu}$ and
$\tilde{F}^{\alpha\mu}$ is the dual of the strength tensor. It is interesting to notice that the coefficients of the quantum breaking $\frac{e^3}{24\pi^2}$, 
$\frac{1}{12\pi^2}\bar{c}_{\mu\nu}$ and $\frac{1}{12\pi^2}b_{\alpha}$ correspond to the beta functions $\frac{\beta_e}{2e}$, $\frac{1}{2}(\beta_{c})_{\mu\nu}$ and $\frac{1}{2}(\beta_{AF})_{\alpha}$, 
respectively \cite{OneLoopLV}, doing the  substitution $b^{\alpha}\rightarrow(k_{AF})^{\alpha}$ that is equivalent to a field redefinition in the latter.

\section{Conclusions} 
\label{s5}

The results show that dimensional Lorentz and CPT-violating coefficients break scale symmetry at the classical level as expected. Since the renormalization process introduces a
scale, we also have in the massless limit the quantum breaking of the energy-momentum tensor trace due to dimensionless coefficients when quantum corrections are taken into 
account. We would also expect that other dimensionless fermion coefficients like $g^{\mu\nu\lambda}$ break scale symmetry in the massive limit because it has an observable piece, the anti-symmetric one in the first two indices.

As a next step, it would be interesting to compute the trace of the energy-momentum tensor for massive fields and considering the
non-minimal SME. The latter framework has a bunch of possibilities in the way it breaks scale symmetry since each non-minimal operator has 
its own scale dimension.

\section{Appendix A} 
\label{A1}

In classical electrodynamics, the canonical energy momentum tensor is 
\bq
&\Theta^{\mu\nu}=\frac{\partial \mathcal{L}}{\partial(\partial_{\mu}\psi)}\partial^{\nu}\psi+\frac{\partial \mathcal
{L}}{\partial(\partial_{\mu}\bar{\psi})}\partial^{\nu}\bar{\psi}+\frac{\partial \mathcal{L}}{\partial(\partial_{\mu
}A_{\alpha})}\partial^{\nu}A_{\alpha}-\eta^{\mu\nu}\mathcal{L}=\nonumber\\
&=i\bar{\psi}\gamma^{\mu}\partial^{\nu}\psi+F^{\lambda\mu}\partial^{\nu}A_{\lambda}-\eta^{\mu\nu}\mathcal{L}.
\label{EMT}
\eq

Although this canonical tensor is not symmetric, it can be symmetrized. The symmetric version it is known as the Belinfante tensor:

\be
T^{\mu\nu}= \Theta^{\mu\nu}+\partial_{\rho}f^{\mu\nu\rho}_A+\partial_{\rho}f^{\mu\nu\rho}_{\psi},
\ee
where $f^{\mu\nu\rho}_{\Phi}=\frac{1}{2}(\Pi^{\rho}S^{\mu\nu}\Phi-\Pi^{\mu}S^{\rho\nu}\Phi+\Pi^{\nu}S^{\mu\rho}\Phi)$ and $S^{\mu\nu}$ is the spin matrix associated with 
the field $\Phi$.

Computing $\partial_{\rho}f^{\mu\nu\rho}$ for the spinor and the vector fields, we find
\bq
&\partial_{\rho}f^{\mu\nu\rho}_{A}=e\bar{\psi}\gamma^{\mu}\psi A^{\nu}+F^{\mu\rho}\partial_{\rho}A^{\nu}, \nonumber\\
&\partial_{\rho}f^{\mu\nu\rho}_{\psi}=\frac{i}{4}(\bar{\psi}\gamma^{\nu}\partial^{\mu}\psi-(\partial^{\mu}\bar{\psi})\gamma^{\nu}\psi-3\bar{\psi}\gamma^{\mu}\partial^{\nu}\psi-(\partial^{\nu}\bar{\psi})\gamma^{\mu}\psi) +\nonumber\\
&+\frac{1}{2}e\bar{\psi}(\gamma^{\nu}A^{\mu}-\gamma^{\mu}A^{\nu})\psi.
\label{eqf}
\eq

Eqs. (\ref{eqf}) leads to the symmetric form of the energy momentum tensor
\bq
T^{\mu\nu}&=\frac{i}{4}[\bar{\psi}\gamma^{\mu}\partial^{\nu}\psi-(\partial^{\mu}\bar{\psi})\gamma^{\nu}\psi+\bar{\psi}\gamma^{\mu}\partial^{\nu}\psi-(\partial^{\nu}\bar{\psi})\gamma^{\mu}\psi]+ \nonumber\\ 
&+F^{\lambda\mu}F_{\lambda}^{\nu}+\frac{1}{2}e\bar{\psi}(\gamma^{\nu}A^{\mu}+\gamma^{\mu}A^{\nu})\psi-\eta^{\mu\nu}\mathcal{L},
\eq
whose trace is $T^{\mu\nu}=m\bar{\psi}\psi$.

In the presence of Lorentz and CPT violation, the energy-momentum tensor is not symmetric under the change of indices as we can see for the pure photon sector \cite{SME}. The
symmetry under Lorentz indices is broken by the Chern-Simons-like term, which also breaks scale symmetry $T^{\mu}_{\mu}=(k_{AF})_{\mu}A_{\alpha}\tilde{F}^{\alpha\mu}$, where
$\tilde{F}^{\alpha\mu}$ is the dual of the tensor $F^{\alpha\mu}$. Considering the fermion sector and the Lagrangian $\mathcal{L}= \frac{1}{2}i \bar{\psi}\Gamma^{\mu}\overleftrightarrow{D}_{\mu}\psi- \bar{\psi}M\psi-\frac{1}{4}F^{\mu\nu}F_{\mu\nu}$, the symmetric energy momentum tensor is given by
\bq
T^{\mu\nu}&=\frac{i}{4}[\bar{\psi}\Gamma^{\mu}\partial^{\nu}\psi-(\partial^{\mu}\bar{\psi})\Gamma^{\nu}\psi+\bar{\psi}\Gamma^{\mu}\partial^{\nu}\psi-(\partial^{\nu}\bar{\psi})\Gamma^{\mu}\psi]+ \nonumber\\ 
&+F^{\lambda\mu}F_{\lambda}^{\nu}+\frac{1}{2}e\bar{\psi}(\gamma^{\nu}A^{\mu}+\gamma^{\mu}A^{\nu})\psi-\eta^{\mu\nu}\mathcal{L},
\eq
whose trace is

\be
T^{\mu}_{\mu}=-\frac{7}{2}i\bar{\psi}\Gamma^{\mu}\partial_{\mu}\psi-\frac{1}{2}i(\partial_{\mu}\bar{\psi})\Gamma^{\mu}\psi-3e \bar{\psi}\Gamma^{\mu}A_{\mu}\psi
+4\bar{\psi}M\psi.
\label{traco}
\ee

Using in eq.(\ref{traco}) the Euler-Lagrange equations, $i\Gamma^{\mu}\partial_{\mu}\psi=-e\Gamma^{\mu}A_{\mu}\psi+M\psi$ and $i(\partial_{\mu}\bar{\psi})\Gamma^{\mu}=e\bar{\psi}\Gamma^{\mu}A_{\mu}-\bar{\psi}M$,  we find out the classical breaking of the dilatation current
for the fermion sector:
\be
T^{\mu}_{\mu}=\bar{\psi}M\psi.
\ee

\section{Appendix B} 
\label{A2}

The result of all finite and regularized divergent integrals for $\Delta^{\mu\nu}$ and $\Pi^{\mu\nu}$ amplitudes of section \ref{s4} is listed below:

\begin{align}
\centering
&\int_k \frac{1}{[(k+p)^2-m^2]^3}= -\frac{\tilde{b}}{6m^2},\\
&\int_k \frac{k^{\mu}}{[(k+p)^2-m^2]^3}= \frac{\tilde{b}}{6m^2}p^{\mu} ,\\
&\int_k \frac{1}{(k^2-m^2)[(k+p)^2-m^2]^2}=-\tilde{b} \iota_0 ,\\
&\int_k \frac{k^{\mu}}{(k^2-m^2)^2[(k+p)^2-m^2]}=\tilde{b} p^{\mu} \iota_1 ,\\
&\int_k \frac{1}{(k^2-m^2)[(k+p)^2-m^2]^3}=\frac{1}{2} \tilde{b} \int_0^1 \frac{x^2}{\D^2} dx, \\
&\int_k \frac{k^{\mu}}{(k^2-m^2)[(k+p)^2-m^2]^3}=\frac{1}{2} \tilde{b} p^{\mu} \int_0^1 \frac{x^3}{\D^2} dx, \\
&\int_k \frac{k^{\mu}}{(k^2-m^2)^2[(k+p)^2-m^2]^2}=\tilde{b} p^{\mu } \int_0^1 \frac{x^2 (1-x)}{\D ^2} dx,\\
&\int_k \frac{k^{\mu}k^{\nu}}{(k^2-m^2)^2[(k+p)^2-m^2]^2}=\tilde{b} \left(p^{\mu } p^{\nu } \int_0^1 \frac{x^3 (1-x)}{\D^2}dx-\frac{1}{2}\eta^{\mu\nu} \int_0^1 \frac{x(1-x)}{\D} dx\right),\\
&\int_k \frac{k^{\mu}k^{\nu}}{(k^2-m^2)[(k+p)^2-m^2]^3}= \frac{1}{2}\tilde{b}\left( p^{\mu} p^{\nu} \int_0^1 \frac{x^4}{\D^2} dx-\frac{1}{2} \eta_{\mu \nu} \int_0^1 \frac{x^2}{\D} dx\right) ,\\
&\int_k \frac{k^{\mu}k^{\nu}}{(k^2-m^2)^2[(k+p)^2-m^2]}=\frac{1}{4}\eta^{\mu\nu}(I_{log}(m^2)-\upsilon_0)-\frac{1}{2}\tilde{b} \eta^{\mu\nu}(Z_0-Z_1)-\tilde{b} p^{\mu}p^{\nu}\iota_2 ,\\
&\int_k \frac{k^{\mu}}{(k^2-m^2)[(k+p)^2-m^2]^2}=-\tilde{b} p^{\mu} (\iota_1-\iota_0 ),\\
&\int_k \frac{k^{\mu}k^{\nu}}{(k^2-m^2)[(k+p)^2-m^2]^2}=\frac{1}{4}\eta_{\mu \nu}(I_{log}(m^2)-\upsilon_0)-\frac{1}{2} \eta_{\mu \nu} \tilde{b}(Z_0-Z_1)-\nonumber\\
&-\tilde{b}(\iota _0-2 \iota _1+\iota _2)p^{\mu} p^{\nu},\\
&\int_k \frac{1}{[(k+p)^2-m^2]}=I_{quad}(m^2)-p^2\upsilon_0, \\
&\int_k \frac{k^{\mu}}{[(k+p)^2-m^2]^2}=-p^{\mu}(I_{log}(m^2)-\upsilon_0), \\
&\int_k \frac{k^{\mu}k^{\nu}}{[(k+p)^2-m^2]^2}=\frac{1}{2}\eta^{\mu \nu}(I_{quad}(m^2)-\upsilon_2 )+p^{\mu}p^{\nu}(I_{log}(m^2)-\xi_0 )-\frac{1}{2}p^2\eta^{\mu\nu}(\xi_0-\upsilon_0),
\end{align}

\begin{align}
\centering
&\int_k \frac{k^{\mu}}{(k^2-m^2)[(k+p)^2-m^2]}=-\frac{1}{2}p^{\mu}(I_{log}(m^2)-\upsilon_0 -\tilde{b} Z_0), \\
&\int_k \frac{k^{\mu}k^{\nu}}{(k^2-m^2)[(k+p)^2-m^2]}=\frac{1}{2}\eta^{\mu\nu}(I_{quad}(m^2)-\upsilon_2 )-\frac{1}{6}\xi_0(p^2\eta^{\mu\nu}+2p^{\mu}p^{\nu})+\nonumber\\
&+\left(\frac{1}{3}p^{\mu}p^{\nu}-\frac{1}{12}p^{2}\eta^{\mu \nu} \right)( I_{log}(m^2)
-\tilde{b} Z_0)+\frac{1}{4}\eta^{\mu\nu}p^2 \upsilon_0-\frac{1}{3}\tilde{b}(p^2\eta^{\mu\nu}-p^{\mu}p^{\nu})\left(\frac{m^2}{p^2}Z_0 +\frac{1}{6} \right), \\
&\int_k \frac{k^{\mu}k^{\nu}k^{\alpha}}{(k^2-m^2)[(k+p)^2-m^2]^2}=-\frac{1}{6}p^{ \{ \alpha} \eta^{\mu\nu \} }(I_{log}(m^2)-\xi_0)+\frac{1}{2}\tilde{b}p^{\{ \alpha} \eta^{\mu\nu \} } Z_2+\nonumber\\ 
&+\tilde{b} p^{\mu}p^{\nu}p^{\alpha}\left( \iota_2 -\iota_3 -\frac{Z_0}{p^2}\right),\\
&\int_k \frac{k^{\mu}k^{\nu}k^{\alpha}k^{\beta}}{(k^2-m^2)[(k+p)^2-m^2]^2}=\frac{1}{24}p^{\{ \alpha} p^{\beta} \eta^{\mu\nu \} }\left( 3(I_{log}(m^2)-\sigma_0)+\tilde{b} +\frac{6 \tilde{b} m^2}{p^2}Z_0-3
\tilde{b}Z_0\right)+\nonumber\\
&+\frac{1}{144}\eta^{\{\alpha \beta }\eta^{ \mu\nu \}}(18(I_{quad}(m^2)-\xi_2 ) -3I_{log}(m^2)+12\xi_0-9\sigma_0-2\tilde{b}-12\tilde{b}m^2 Z_0+3\tilde{b}Z_0)+\nonumber\\
&+\frac{1}{48}\tilde{b}p^{\alpha}p^{\beta}p^{\mu}p^{\nu}\left(-3\iota_0-\frac{10}{p^2}+\frac{45}{p^2}Z_0-\frac{60m^2}{p^4}Z_0 \right)\\
&\int_k \frac{k^{\mu}k^{\nu}k^{\alpha}}{(k^2-m^2)^2[(k+p)^2-m^2]}=-\frac{1}{12}p^{ \{ \alpha} \eta^{\mu\nu \} }(I_{log}(m^2)-\xi_0- 6\tilde{b}Z_1+6\tilde{b}Z_2)+\tilde{b}p^{\mu}p^{\nu}p^{\alpha}\iota_3,\\
&\int_k \frac{k^{\mu}k^{\nu}k^{\alpha}k^{\beta}}{(k^2-m^2)^2[(k+p)^2-m^2]}=\frac{1}{24}\eta^{\{\alpha \beta }\eta^{ \mu\nu \}}(3(I_{quad}(m^2)-
\xi_2)- p^2(I_{log}(m^2)-\xi_0 ) ) +\nonumber\\
&+\frac{1}{48}(p^2\eta^{\{\alpha \beta }\eta^{ \mu\nu \}}+2p^{\{ \alpha} p^{\beta} \eta^{\mu\nu \} })(I_{log}(m^2)-\sigma_0)-\frac{1}{8}\tilde{b}p^2(Z_0-5Z_2+2Z_3)\eta^{\{\alpha \beta }\eta^{ \mu\nu \}}+\nonumber\\
&+\frac{1}{2}\tilde{b}p^{\{ \alpha} p^{\beta} \eta^{\mu\nu \} }(Z_3-Z_2)-\tilde{b}p^{\alpha}p^{\beta}p^{\mu}p^{\nu}\iota_4,
\end{align}
where $Z_n=Z_n(p^2,m^2)=\int_0^1 dx \ln \frac{\D}{m^2}$, $\iota_n=\int_0^1 dx \frac{x^n(1-x)}{\D}$, $\tilde{b}\equiv \frac{i}{(4\pi)^2}$ and $\D=\D(x)=m^2-p^2 x(1-x)$.

	

\end{document}